# Analysis of thermally stimulated luminescence and conductivity without quasiequilibrium approximation

A. Opanowicz

Technical University of Łódź, Institute of Physics, Łódź , Poland

Thermally stimulated luminescence (TSL) and conductivity (TSC) are considered using the classical insulator model that assumes one kind of the active trap, one kind of inactive deep trap, and one kind of the recombination center. Kinetic equations describing the model are solved numerically without and with the use of the quasiequilibrium (QE) approximation. The QE state parameter $q_l$, the relative recombination probability $\gamma$, and a new parameter called quasi-stationary (QS) state parameter $q^* = q_l \gamma$ are used for the analysis of the TSL and TSC. The TSL and TSC curves and the temperature dependences of $q_l$, $q^*$, $\gamma$, the recombination lifetime, and the occupancies of active traps and recombination centers are numerically calculated for five sets of kinetic parameters and different heating rates. These calculation results show that: (1) the upper limit of the heating rate for presence of the QS state appears at higher heating rate than that for the QE state when the retrapping process is present, and (2) the TSL (TSC) curves in the QS state have the properties similar to those for the TSL (TSC) curves in the QE state. Approximate formulas for calculation of the parameters $q_l$ and $q^*$ in the initial range of the TSL and TSC curves are derived and used in the heating-rate methods, proposed in this work, for determination of those parameters from the calculated TSL curves.

## 1. Introduction

It is well-known that thermally stimulated luminescence (TSL, or thermoluminescence) and thermally stimulated conductivity (TSC) are phenomena that appear when a crystal of insulator or semiconductor previously photoexcited at low temperature is next heated in dark. Generally accepted explanation of the phenomena assumes thermal release of photoexcited charge carrier captured by trap to the delocalized band (TSC) following radiative recombination (TSL) of the carrier with the opposite-sign carrier in the recombination center (e.g.[1,2]). In the pioneering theories of TSL and TSC it was assumed that the thermally released carriers are in quasi-equilibrium (QE) in the delocalized band [3-7]. The assumption makes it possible to derive analytical description of the TSL and TSC curves. Later the QE assumption was often used by many authors (e.g. [8-12]) for description of the TSL and TSC processes. However, numerical solutions of the basic kinetic equations describing the TSL and TSC processes showed that the QE approximation can be questioned. Kelly and colleagues [13] compared the results of numerical solutions of the kinetic equations with the ones obtained from the analytical expressions. They found that the QE approximation is not adequate for the low trap density ($<10^{15}$cm$^{-3}$). Later several more authors carried out the numerical solutions of the kinetic equations and calculated the TSL and TSC characteristics [14-25]. Shenker and Chen [14] found that the QE approximation in description of TSL is true both in the weak and strong retrapping cases and that the level of QE deteriorates in the high temperature part of the TSL peak calculated for the strong retrapping case. Lewandowski and McKeever [15] introduced new parameters $Q(T) = q_l(T) +1$, and $P(T)$ for analysis of the validity of the QE assumption, where $q_l =[dn_c(T)/dt]/I_e(T)$ is the QE level parameter, $P(T)$ is the ratio of the retrapping rate to the recombination rate, $n_c$ is the free electron density, $I_e$ is the TSL intensity, $T$ is the temperature, and $t$ is the time. They calculated the TSL and TSC characteristics without the QE approximation. On the basis of the results they concluded that the use of the QE approximation in description of TSL and TSC in the case of strong retrapping is without merit and therefore the traditional strong retrapping equations should be considered worthless. They also concluded that the slow retrapping (first order kinetic) case satisfies the requirements of a realistic system, and this in turn may explain the apparent dominance of first order kinetics in nature. They suggested an experimental method for determining the shape of the $q_l(T)$ function using the simultaneously measured TSL and TSC curves. The analysis of the TSL and TSC without the QE approximation was developed in next works of Lewandowski and colleagues [16,17]. However, the conclusions of Lewandowski and colleagues [15-17] are not in agreement with the results of numerical solutions of kinetic equations published by Shenker and Chen [14], Opanowicz [18,19], and Opanowicz and Przybyszewski [20]. In works [19,20] the TSL and TSC characteristics were calculated without the use of the QE approximation and the results were compared to the corresponding results obtained using the QE approximation. The results are as follows: (1) the QE approximation is correct with accuracy 1% when the density of the recombination centers and traps is higher than $10^{14}$ cm$^{-3}$ and the recombination coefficient is higher than $10^{-12}$cm$^3$ s$^{-1}$ , (2) the QE assumption is valid for the strong retrapping case, and (3) the TSL curve calculated for the strong retrapping case can have the first-order shape. These results have been later confirmed by the results of Sunta and colleagues that analysed the TSL processes using different models [22-25]. Sunta and colleagues [24] found also that the level of QE during the TSL emission depends on the heating rate of a sample, and that near the turning point from the QE to non-QE state the TSL begins to change its shape. They used this property for derivation of the experimental method to test whether or not the TSL emission is occurring under QE state.

In the present work we report the TSL and TSC curves calculated numerically (with and without the use of the QE approximation) for an insulator model with one kind of recombination centers and two kinds of traps (active shallow and inactive deep). The curves calculated at different heating rates are analyzed to understand the TSL and TSC processes when the QE conditions are not fulfilled. In order to attain it we calculated and discussed other characteristics of the TSL and TSC: the temperature dependences of the recombination probability, the recombination lifetime, the occupancy of active traps, the QE parameter $q_I$ , and the thermal generation rate from the active traps. A new parameter $q^* =[dn_c/dt/(I_e+I_t)]$ ($I_t$ is the trapping intensity) is introduced, it takes into account the participation of both the recombination and retrapping in producing the quasi-stationary (QS) state of the TSL and TSC. The methods for determination of the QS level from the TSL and TSC curves are proposed and examined. There are also reported the results of determination of the trap depth from the non-QE TSL curves with four literature methods (Garlick and Gibson [4], Hoogenstraaten [5], Haering and Adams [6]), and Chen[2].

## 2. Theoretical description of the TSC and TSL without quasi-equilibrium approximation

In order to describe the TSC and TSL we will consider an n-type insulator model with one kind of the recombination center and two kinds of the electron traps (e.g.[9-12]). At low temperature the recombination centers and traps are empty (ionized). At the temperature the insulator is subjected to external photo-excitation (with photon energy greater than the band gap) that produces free electrons and holes. The holes are captured by the recombination centers while the electrons are captured by different kinds of traps. It is assumed that there are the shallow (active) traps that are partly or fully filled with electrons during photo-excitation, and the thermally disconnected (inactive) deep traps that are fully filled by electrons. After removing of excitation the insulator is heated in the dark according to linear heating rate $\omega=dT/dt=Const(T)$. The heating provides thermal energy to the captured electrons in the active traps that are excited to the conduction band giving increase in the conductivity (TSC). The excited electrons recombine radiatively (TSL) in the recombination centres or are retrapped by empty active traps. The electrons in deep traps stay immobilized. According to the model the TSC and TSL processes can be described by the neutrality condition

$$n_c + n_t + M = p_r, \qquad (1)$$

and the kinetic equations

$$\frac{dn_c}{dt} = n_t s \exp\left(-\frac{E}{kT}\right) - n_c \alpha\, p_r - n_c \beta(N_t - n_t), \qquad (2)$$

$$\frac{dn_t}{dt} = n_c \beta(N_t - n_t) - n_t s \exp\left(-\frac{E}{kT}\right), \qquad (3)$$

$$\frac{dn_c}{dt} + \frac{dn_t}{dt} = \frac{dp_r}{dt} \; , \tag{4}$$

where $n_c$ is the density of free (in the conduction band) electrons representing TSC, $n_t$ is the density of electrons captured by the active traps with the density $N_t$, $M$ is the density of electrons in the inactive deep (thermally disconnected) traps, $\alpha$ is the recombination coefficient for the free electron with the hole in the recombination center, $p_r$ is the density of holes in the recombination centers, $s$ is the frequency factor for the active trap, and $\beta$ is the trapping coefficient for the free electron by the active trap. It is assumed that $\alpha$, $\beta$, and $s$ are independent of the temperature, and the luminous efficiency is equal to one.

The TSL intensity is represented by equations

$$I_e = -\frac{dp_r}{dt} = n_c \alpha \; p_r = \frac{n_c}{\tau_e} \; , \tag{5}$$

where

$$\tau_e = \frac{1}{\alpha \; p_r} \tag{6}$$

is the free electron recombination lifetime.

The equation system (1-4) can be solved assuming that free electrons are in the quasiequilibrium (QE) state with the recombination centers that is mathematically expressed by condition (e.g.[9-11])

$$\left|\frac{dn_c}{dt}\right| << \left|\frac{dn_t}{dt}\right| \approx \left|\frac{dp_r}{dt}\right| . \tag{7}$$

Additionally, it is assumed that (e.g.[9,15])

$$n_c << n_t + M \; . \tag{8}$$

The approximations (7) and (8) are called the quasiequilibrium (QE) approximation. We will consider the TSL and TSC processes without the QE approximation applying the QE parameter [15]:

$$q_I = \frac{\frac{dn_c}{dt}}{I_e} \; . \tag{9}$$

Using the QE parameter in equation (2) we can derive the equation for the TSC intensity in the form :

$$n_c = \frac{n_t \, s \exp\left(-\frac{E}{kT}\right)}{\alpha \, p_r (q_I + 1) + \beta (N_t - n_t)} \; . \tag{10}$$

Putting expression (10) for the TSC intensity into equation (5) we have:

$$I_e = \frac{\alpha \, p_r \, n_t \, s \exp\left(-\frac{E}{kT}\right)}{\alpha \, p_r (q_I + 1) + \beta (N_t - n_t)} . \tag{11}$$

For the description of the TSL and TSC we can use the recombination probability parameter [11,12]

$$\gamma = \frac{\alpha \, p_r}{\alpha \, p_r + \beta (N_t - n_t)} = \frac{I_e}{I_e + I_t} , \tag{12}$$

where

$$I_t = n_c \beta (N_t - n_t) \tag{13}$$

is the intensity of free electron trapping. Taking into account that

$$\tau_t = \frac{1}{\beta (N_t - n_t)} \tag{14}$$

is the free electron trapping time we can present the parameter $\gamma$ in the form:

$$\gamma = \frac{\tau_e^{-1}}{\tau_e^{-1} + \tau_t^{-1}} = \frac{P_e}{P_e + P_t} \tag{15}$$

where $P_e = \tau_e^{-1}$ and $P_t = \tau_t^{-1}$ are the probabilities (per second) of the recombination and, respectively, trapping. The parameter $\gamma$ can have values from 0 to 1. $0.5 < \gamma < 1$ is for the weak-retrapping case, and $\gamma < 0.5$ corresponds to the strong retrapping case. Now using the parameter $\gamma$ we can present equation (11) in the simpler form

$$I_e = \frac{\gamma G}{\gamma q_I + 1}, \tag{16}$$

where

$$G = n_t s \exp\left(-\frac{E}{kT}\right) \tag{17}$$

is the thermal excitation rate of electrons from the active (shallow) traps to the conduction band. When $|q_I| << 1$ equation (16) takes the form typical for the QE state [11,12]

$$I_e = \gamma G . \tag{18}$$

However, when there is non-QE state, equation (16) may take also the form (18) because the condition $|\gamma q_I| << 1$ can be realized for strong retrapping (i.e. $\gamma << 1$).
The condition for the QE state can be also expressed in the form

$$\left|\frac{dn_c}{dt}\right| << \left|\frac{dn_t}{dt} = I_t - G\right|, \tag{19}$$

where $dn_t/dt < 0$ because $G > I_t$. So, it follows from inequality (19) that the retrapping process counteracts reaching of the QE state. On the other hand, formula (9) transformed (using relations $I_e = n_c/\tau_e$ (equation.(5)) and $\omega = dT/dt$) into the form

$$q_I = \frac{\omega \tau_e}{n_c} \frac{dn_c}{dT} \tag{20}$$

suggests that $\tau_e$ strongly influences the QE level. Using the parameters $\tau_e$, $\gamma$, $q_I$, and $G$ equation (10) for the TSC intensity can be presented in the form corresponding to equation (16) for the TSL intensity:

$$n_c = \frac{\tau_e \gamma G}{\gamma q_I + 1} . \tag{21}$$

Let us to introduce a new parameter

$$q^* = \gamma q_I . \tag{22}$$

Putting the formulas for $\gamma$ (12), $q_I$ (9), $I_e$ (11), and $I_t$ (13) into the equation (22) we obtain

$$q^* = \frac{\dfrac{dn_c}{dt}}{n_c \alpha p_r + n_c \beta (N_t - n_t)} = \frac{\dfrac{dn_c}{dt}}{I_e + I_t} . \tag{23}$$

From equation (23) we see that the parameter $q^*$ describes the state of the free electrons in relation to the sum of the recombination and trapping processes. The state determined by condition

$$|q^*| << 1 \tag{24}$$

we call the quasi-stationary state (QS) because the state corresponds to $dn_c/dt \approx 0$ as it follows from equation (2). The stationary state of free electrons appears at the TSC peak-maximum position where $dn_c/dt = 0$. Of course, from equation (22) it is clear that $q^* \lesssim q_I$ because $\gamma \lesssim 1$.

## 3. Methods for determination of the parameters of the QE and QS states

In this section we will consider the TSC and TSL in their initial ranges. Using formulas (17) and (22) and a new parameter $R = \gamma n_t/N_t$ we can present formula (21) for the TSC intensity in the form

$$n_c(T) = \frac{\tau_e R N_t s \exp\left(-\frac{E}{kT}\right)}{(q^* + 1)kT^2}. \tag{25}$$

We assume that $R$, $\tau_e$, and $q^*$ are independent of the temperature in the initial range of TSL, i.e. for $I(T)/I_{em}<0.5$ where $I_{em} = I_e(T_{ml})$ is the TSL peak-maximum intensity. In this case differentiation of equation (25) gives

$$\frac{dn_c}{dT} = \frac{\tau_e R N_t s E \exp\left(-\frac{E}{kT}\right)}{(q^* + 1)kT^2}. \tag{26}$$

Setting $n_c$ and $dn_c/dT$, given respectively by formulas (25) and (26), into equation (20) we obtain

$$\frac{q_I}{\tau_e} = \frac{\omega E}{kT^2}. \tag{27}$$

This simple formula makes it possible to calculate the ratio $q_I/\tau_e$ when the values of $E$, $T$, and $\omega$ are known. When the value of $\tau_e$ is known we can also calculate the value of $q_I$ with this formula. Using formula (27) for two equal values of the relative TSL (or TSC) intensity corresponding to two different values of $\omega$, and assuming $\tau_e = \text{Const}(\omega)$ one can derive formula describing variation of the parameter $q_I$ with $\omega$ in the initial range of TSL (or TSC):

$$\frac{q_{I2}}{q_{I1}} = \frac{\omega_2}{\omega_1} \frac{T_1^2}{T_2^2}, \tag{28}$$

where $q_{I1} = q_I[T(\omega_1)] = q_I(T_1)$, and $q_{I2} = q_I[T(\omega_2)] = q_I(T_2)$.

In order to derive the formula for determination of the QS parameter $q^*$ we use equation (16) for the TSL intensity in the form corresponding to equation (25) for the TSC:

$$I_e(T) = \frac{R N_t s \exp\left(-\frac{E}{kT}\right)}{q^* + 1}. \tag{29}$$

Using this equation we can find the ratio of the TSL intensities $I_{e1}(\omega_1)$ and $I_{e2}(\omega_2)$ at two different heating rates $\omega_1$ and $\omega_2 > (\omega_1)$

$$\frac{I_{e1}}{I_{e2}} = \frac{R_1 (q_2^* + 1) \exp\left(-\frac{E}{kT_1}\right)}{R_2 (q_1^* + 1) \exp\left(-\frac{E}{kT_2}\right)}. \tag{30}$$

Assuming that $I_{e1}$, $I_{e2}$, and $R_1(\omega_1) = R_2(\omega_2)$ correspond to equal value of the relative TSL intensities at two heating rates (i.e. $I_e(T_1(\omega_1))/I_{em1} = I_e(T_2(\omega_2))/I_{em2}$) in the initial range of TSL we obtain

$$\frac{I_{e1}}{I_{e2}} = \frac{(q_2^* + 1) \exp\left(-\frac{E}{kT_1}\right)}{(q_1^* + 1) \exp\left(-\frac{E}{kT_2}\right)}. \tag{31}$$

Approximate equations (28) and (31) are the system with two unknowns $q_1^*$ and $q_2^*$. The solution of the system is:

$$q_1^* = \frac{Z - 1}{Y - Z}, \tag{32}$$

$$q_2^* = q_1^* Y, \tag{33}$$

where

$$Y = \frac{\omega_2 T_1^2}{\omega_1 T_2^2}, \tag{34}$$

$$Z = \frac{I_{e1} \exp\left(-\frac{E}{kT_2}\right)}{I_{e2} \exp\left(-\frac{E}{kT_1}\right)}. \tag{35}$$

Formulas (32), (33), (34), and (35) may be used for determination of the values of $q_1^*$ and $q_2^*$ when the values of $T_1$, $T_2$, $I_{e1}$, and $I_{e2}$ (for $I_{e1}/I_{em1} = I_{e2}/I_{em2}$), and the value of $E$ are known from the initial ranges of the TSL curves measured at two heating rates $\omega_1$ and $\omega_2$.

Another method for determination of $q^*$ in the initial range of the TSL curve can be derived assuming that the slower heating rate $\omega_1 (=\omega_s)$ is sufficient to obtain the QS state, i.e. $|q_1^*| \ll 1$. In this case equation (31) can be simplified and transformed to the form

$$q^*(T_2) = \frac{I_{e1} \exp\left(-\frac{E}{kT_2}\right)}{I_{e2} \exp\left(-\frac{E}{kT_1}\right)} - 1. \tag{36}$$

For determination of $q^*$ with equation (36) we need the value of $E$ and the values of $T_1$, $T_2$, $I_{e1}$, and $I_{e2}$ corresponding to equal values of the relative TSL intensity of two TSL curves measured at $\omega_1 = \omega_s$ and $\omega_2 > \omega_1$.

## 4. Calculation results of the TSL and TSC characteristics

In order to analyze the QE and QS problems we numerically calculated the temperature dependences of different TSL and TSC characteristics: $I_e(T)$, $n_c(T)$, $n_t(T)$, $p_r(T)$, $dn_c(T)/dt$, $\chi(T)$, $\tau_e(T)$, $\gamma(T)$, $q_t(T)$, $q^*(T)$, and the TSL (TSC) symmetry factor $\mu_g = (T_2 - T_m)/(T_2 - T_1)$ [9], where $T_m$ is the TSL (TSC) peak–maximum temperature, and $T_1$ and $T_2(>T_1)$ are the temperatures at the intensity of TSL (TSC) equal to 0.5 of the peak–maximum intensity. For the calculations we solved numerically (without the QE approximation (NQE)) the kinetic equations (1)-(4) using the Runge-Kutta [26] and Gear [27] procedures and Pentium computer. The TSL and TSC characteristics were also calculated using equations (1)-(4) that are approximated under QE conditions (7) and (8) and the method described in our previous paper [11]. The calculations were performed for values of the sample heating rate in the range $\omega = 10^{-3} - 10$ K s$^{-1}$. For discussion we present example results of the TSL and TSC characteristics calculated for five sets of traps and recombination centers parameters listed in table 1.

**Table 1**. Kinetic parameters of traps used in calculations: The density of active traps ($N_t$), the initial density of electrons in the active traps ($n_{t0}$), the trapping coefficient ($\beta$), the frequency factor ($s$), the density of the deep inactive traps ($M$), the recombination coefficient ($\alpha$), the initial density of holes in the recombination centres $p_{r0} = M + n_{t0}$, and the active trap depth $E = 0.3$ eV.

| Set | $N_t$(cm$^{-3}$) | $n_{t0}$(cm$^{-3}$) | $\beta$(cm$^3$s$^{-1}$) | $s$(s$^{-1}$) | $M$(cm$^{-3}$) | $\alpha$(cm$^3$s$^{-1}$) |
|-----|---|---|---|---|---|---|
| A | $10^{11}$ | $10^{11}$ | $10^{-13}$ | $10^7$ | $10^8$ | $10^{-12}$ |
| B | $10^{12}$ | $10^{11}$ | $10^{-12}$ | $10^7$ | $10^9$ | $10^{-12}$ |
| C | $10^{11}$ | $10^{10}$ | $10^{-12}$ | $10^7$ | $10^{12}$ | $10^{-13}$ |
| C$_1$ | $10^{11}$ | $10^{10}$ | $10^{-12}$ | $10^7$ | $5 \times 10^{10}$ | $3 \times 10^{-13}$ |
| D | $10^{11}$ | $10^{10}$ | $10^{-13}$ | $10^6$ | $10^{11}$ | $10^{-12}$ |

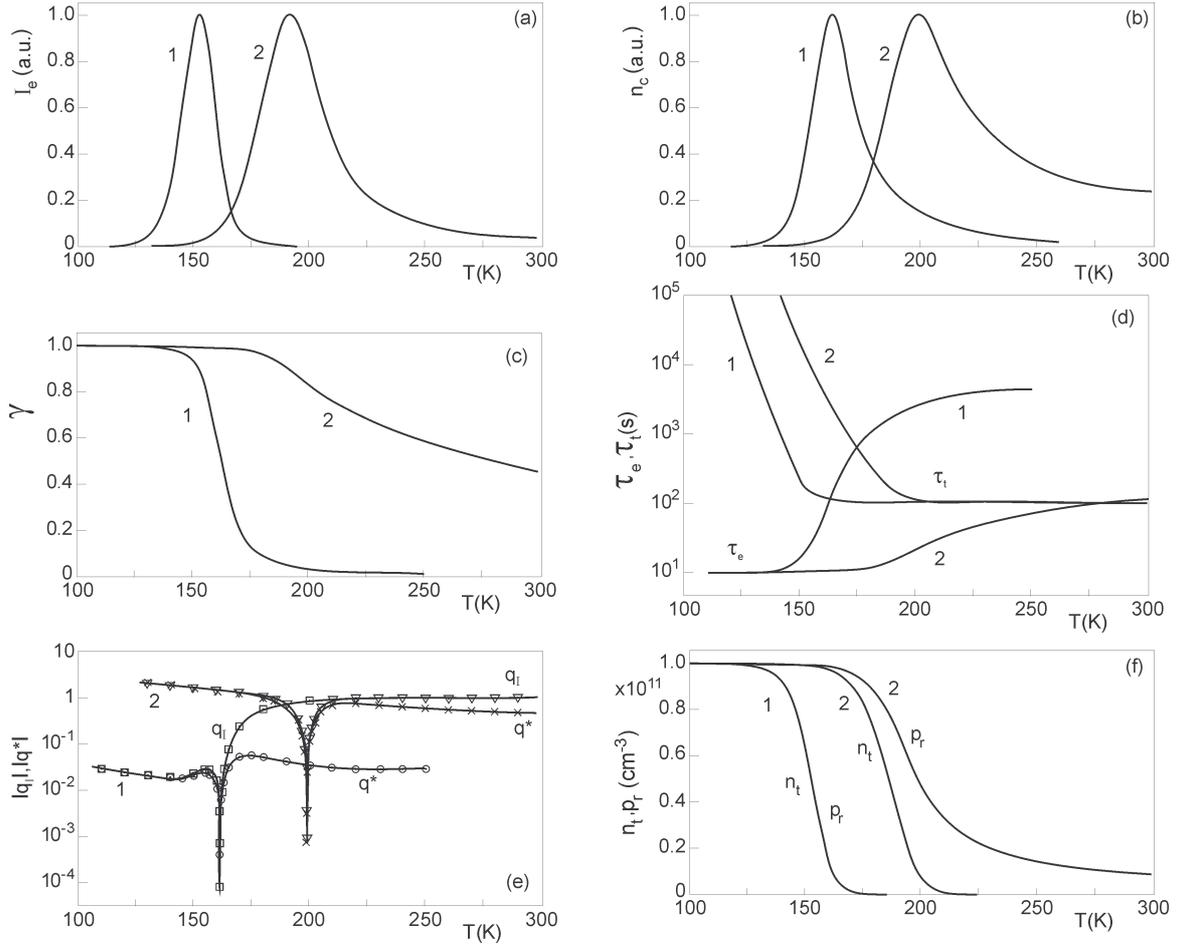

**Figure 1**. Temperature dependences of TSL and TSC characteristics calculated for set A of kinetic parameters and for two values of the heating rate ($\omega = 10^{-2}$ (1) and 1 (2) Ks$^{-1}$): (a) the relative TSL intensity ($I_e(T)$) (a.u. means arbitrary units), (b) the relative TSC intensity ($n_c(T)$), (c) the relative recombination probability ($\gamma(T)$), (d) the recombination lifetime ($\tau_e(T)$), (e) the QE ($q_I(T)$) and QS ($q^*(T)$) parameters ($q_I > 0$ and $q^* > 0$ at $T < T_{mn}$, $q_I < 0$ and $q^* < 0$ at $T > T_{mn}$, and $q_I = q^* = 0$ at $T_{mn}$), and (f) the densities of electrons in traps ($n_c(T)$) and holes in recombination centers ($p_r(T)$).

*4.1. Case A*
Figures 1a to 1d show the TSL and TSC characteristics numerically calculated without the QE approximation for Set A of kinetic parameters (weak retrapping case), and two values of heating rates $\omega = 10^{-2}$ (1) and 1 Ks$^{-1}$ (2). It is seen from figures 1a and 1b that the TSL and both TSC peaks calculated for $\omega = 10^{-2}$ K s$^{-1}$ have the shape similar to that typical for the corresponding TSL and TSC curves calculated for the set A using the QE approximation in the case of the weak retrapping (see e.g.[2,11,12], and table 2). The following figures show that it is the case of decreasing the recombination probability ($\gamma(T)$, Fig.1c) and increasing the recombination lifetime ($\tau_e(T)$, Fig.1d) with increasing of the temperature. Analysis of the temperature dependence of the QE level parameter $q_I$ for $\omega = 10^{-2}$ Ks$^{-1}$ ((Fig. 1e) suggests that the TSL and TSC occur in approximate QE state ($|q_I| < 1.7 \times 10^{-2}$) except for high temperatures ($T > 160$ K). On the other hand the values of the QS parameter ($|q^*| = |\gamma q_I| = 1.1 \times 10^{-2}$ at $T < 160$ K, and $|q^*| < 5 \times 10^{-2}$ at $T > 160$ K) show that the QS state is may be assumed to be present in the whole interesting temperature range. This is because of the increasing retrapping probability with increasing temperature ($\gamma < 0.5$ at $T > 165$ K). The TSL characteristics for the set A of kinetics parameters were also calculated using the QE approximation (table 2). The satisfactory agreement between the values of the parameters supports the above-drawn conclusions concerning the presence of the approximate QE state during the TSL and TSL processes in Case A at $\omega = 10^{-2}$ Ks$^{-1}$.

**Table 2.** Values of parameters of TSL and TSC curves ($T_{m(n)}$, $\mu_{g(n)}=(T_{2I(n)}-T_{mI(n)})/(T_{2I(n)}-T_{1I(n)})$, $I_{em}$, and $n_{cm}$) calculated for set A of kinetic parameters without the QE approximation (NQE) and with the QE approximation (QE) and for two values of heating rate $\omega$. $T_{1I(n)}$ and $T_{2I(n)}$ ($>T_{1I(n)}$) are the temperatures at the TSL(TSC) intensity equal to 0.5 of the peak-maximum intensity $I_{em}(n_{cm})$ at $T_{mI(n)}$.

| TSL | $\omega$(Ks$^{-1}$) | $T_{mI}$(K) | $T_{1I}$ (K) | $T_{2I}$(K) | $\mu_{gI}$ | $I_{em}$(cm$^{-3}$s$^{-1}$) |
|---|---|---|---|---|---|---|
| NQE | 10$^{-2}$ | 153.4 | 144.3 | 160.8 | 4.48x10$^{-1}$ | 5.31x10$^{7}$ |
| QE |  | 153.4 | 144.3 | 160.6 | 4.43x10$^{-1}$ | 5.43x10$^{7}$ |
| NQE | 1 | 192.5 | 177.8 | 208.7 | 5.26x10$^{-1}$ | 2.27x10$^{9}$ |
| QE |  | 188.2 | 174.5 | 199.0 | 4.47x10$^{-1}$ | 3.66x10$^{9}$ |

| TSC | $\omega$(Ks$^{-1}$) | $T_{mn}$(K) | $T_{1n}$(K) | $T_{2n}$(K) | $\mu_{gn}$ | $n_{cm}$(cm$^{-3}$) |
|---|---|---|---|---|---|---|
| NQE | 10$^{-2}$ | 161.5 | 152.1 | 174.4 | 5.81x10$^{-1}$ | 2.00x10$^{9}$ |
|  | 1 | 198.7 | 182.9 | 229.3 | 6.58x10$^{-1}$ | 4.00x10$^{10}$ |

The TSL and TSC peaks calculated for $\omega$=1 Ks$^{-1}$ are shifted toward the higher temperatures and they are wider in relation to the TSL and TSC peaks calculated for $\omega$=10$^{-2}$ Ks$^{-1}$ (figures 1a and 1b, and table 2). The intensity of the TSL peak at $\omega$=1Ks$^{-1}$ in the non-QE case is lower than that for the QE case and the intensity of the TSL at the higher temperatures ($T$>200 K) are higher for the non-QE case than for the QE case. The results of $q_I(T)$ for $\omega$=1 Ks$^{-1}$ (figure 1e) show that the rate of change of free electrons ($dn_c/dt$) is relatively higher in relation to the TSL intensity ($I_e$) than at $\omega$=10$^{-2}$Ks$^{-1}$ causing that the QE state is not present in this case. It explains the strong deviation of the parameters of the non-QE TSL and TSC curves from those for the QE state. The condition for the QS state (figure 1e) is also not achieved because of the value of $\gamma$=0.5-1. In the high temperature range ($T$=250 – 300 K) of the TSL and TSC the parameters QE and QS reach approximately constant values, $|q_I|\approx1$, and $|q^*|\approx0.5$ (figure 1e). It means that the rate of change of free electron density is approximately equal to the TSL intensity and the trapping intensity, $|dn_c/dt|\approx I_e\approx I_t$.

Fig.1f presents the occupancies of the active traps ($n_t(T)$) and the recombination centers ($p_r(T)$) calculated for the two values of the heating rates ($\omega$=10$^{-2}$ and 1 Ks$^{-1}$) for Case A. At $\omega$=10$^{-2}$ it is observed that $n_t\approx p_r$ in the whole range of the TSL an TSC processes. In the non-QE case ($\omega$=1Ks$^{-1}$) we see that the relation $n_t\approx p_r$ holds only in the initial temperature range of the TSL and TSC. At higher temperatures we have $p_r>n_t$ (T>150 K), and $p_r>>n_t$ (T>200 K), and $p_r$ decreases very slowly with increasing temperature. This result is the consequence of the fact that the recombination intensity falls behind the strong generation rate of free electrons even in low temperature range. Consequently part of the holes accumulated in the recombination centers recombine only at the high temperatures. The process is also represented by the temperature dependences of the relative recombination probability ($\gamma(T)$, Fig.1c) and the recombination lifetime ($\tau_e(T)$, Fig.1d). Different behavior is observed in the case of the temperature dependence of the trapping time ($\tau_t(T)$, Fig.1d) that is similar to that in the QE case. The trapping time is dependent on the density of empty active traps that is determined mainly by the thermal generation rate of free electrons from the traps.

*4.2. Case B*

The TSL and TSC characteristics for Case B calculated without QE approximation for $\omega$=10$^{-2}$ and 1 Ks$^{-1}$ are presented in figures 2a to 2d and table 3. The TSL and TSC curves for $\omega$=10$^{-2}$Ks$^{-1}$ (figures 2a and 2b) have the shape typical for the strong retrapping and the recombination lifetime increasing with the temperature (figure 2c, and [2,11]). The values of the QE parameter ($|q_I|<10^{-2}$ at $T$<190 K), figure 2d) suggest that the QE state is present. This conclusion is confirmed by the results of the TSL parameters calculated for Case B and $\omega$=10$^{-2}$ Ks$^{-1}$ using the QE approximation (table 3). The properties of the TSL peak at $T$<220 K remain typical for the QE state when the heating rate is increased from $\omega$=10$^{-2}$ to 10$^{-1}$Ks$^{-1}$ (table 3). But the QE state is not present for $\omega$= 10$^{-1}$Ks$^{-1}$ because in this case $|q_I|$=1.3x10$^{-1}$($|q^*|$=1.3x10$^{-2}$) at $T$=160K, and $|q_I|$= 2.4x10$^{-1}$($|q^*|$=2.3x10$^{-2}$) at $T$=220 K.

The results of $q^*(T)$ suggest that $\omega_s \approx 10^{-1} Ks^{-1}$ corresponds to the limit heating rate for presence of the QS state while the limit of validity of the QE state is at the slower heating rate, $\omega_q \approx 10^{-2} Ks^{-1}$.

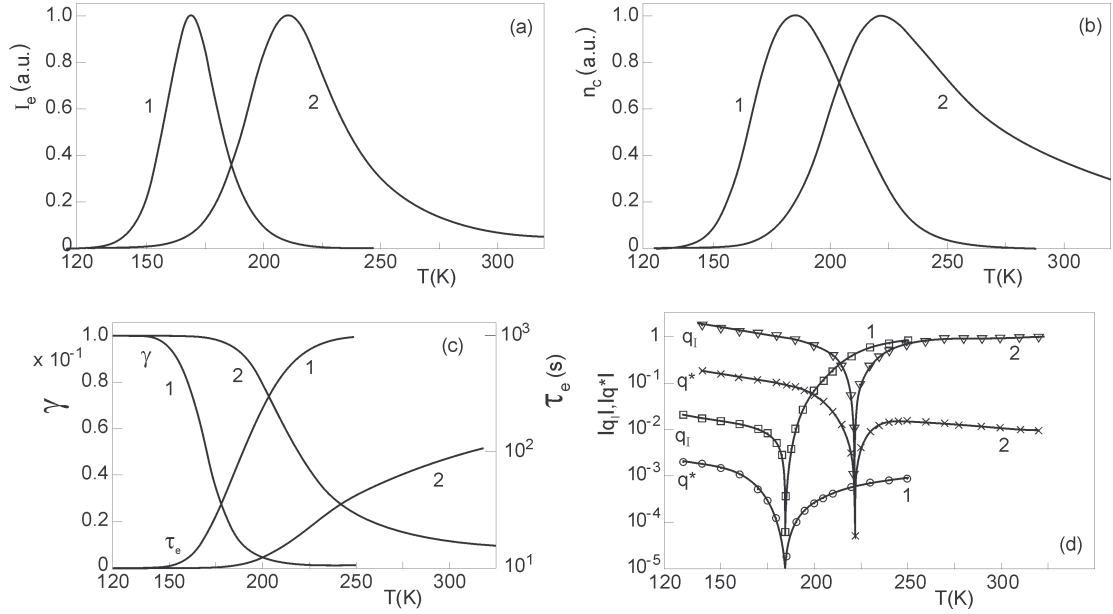

**Figure 2**. Temperature dependences of TSL and TSC characteristics calculated for set B of kinetic parameters and for two values of the heating rate ($\omega = 10^{-2}$ (1) and 1 (2) $Ks^{-1}$): (a) the relative TSL intensity ($I_e(T)$), (b) the relative TSC intensity ($n_c(T)$), (c) the relative recombination probability ($\gamma(T)$) and the recombination lifetime ($\tau_e(T)$), and (d) the parameters $q_l(T)$ and $q^*(T)$.

**Table 3**. Values of parameters of TSL and TSC curves calculated for set B of kinetic parameters without the QE approximation (NQE) and with the QE approximation (QE) and for different values of heating rate $\omega$.

| TSL | $\omega(Ks^{-1})$ | $T_{ml}(K)$ | $T_{1l}(K)$ | $T_{2l}(K)$ | $\mu_{gl}$ | $I_{em}(cm^{-3}s^{-1})$ |
|---|---|---|---|---|---|---|
| NQE | $10^{-2}$ | 168.9 | 156.3 | 182.4 | $5.26 \times 10^{-1}$ | $3.31 \times 10^7$ |
| QE  |           | 168.9 | 156.4 | 182.8 | $5.26 \times 10^{-1}$ | $3.33 \times 10^7$ |
| NQE | $10^{-1}$ | 187.7 | 172.3 | 205.2 | $5.32 \times 10^{-1}$ | $2.59 \times 10^8$ |
| QE  |           | 187.9 | 172.6 | 205.1 | $5.28 \times 10^{-1}$ | $2.71 \times 10^8$ |
| NQE | 1         | 210.3 | 190.7 | 236.4 | $5.71 \times 10^{-1}$ | $1.64 \times 10^9$ |
| QE  |           | 211.4 | 192.3 | 233.0 | $5.31 \times 10^{-1}$ | $2.16 \times 10^9$ |

| TSC | $\omega(Ks^{-1})$ | $T_{mn}(K)$ | $T_{1n}(K)$ | $T_{2n}(K)$ | $\mu_{gn}$ | $n_{cm}(cm^{-3})$ |
|---|---|---|---|---|---|---|
| NQE | $10^{-2}$ | 184.8 | 164.8 | 213.2 | $5.88 \times 10^{-1}$ | $8.96 \times 10^8$ |
|     | 1         | 222.0 | 197.3 | 274.9 | $6.82 \times 10^{-1}$ | $2.83 \times 10^{10}$ |

The TSL and TSC peaks calculated for $\omega=1$ $Ks^{-1}$ (figures 2a and 2b, and table 3) are shifted toward the higher temperatures in relation to the TSL ad TSC curves for $\omega=10^{-2} Ks^{-1}$ and they show "abnormal" broadening at the high temperature range. The TSL and TSC at $\omega=1$ $Ks^{-1}$ appear in the non-QE state as it follows from the calculated results of $q_l(T)$ (figure 2d) and the comparison of the TSL parameters calculated for $\omega=1 Ks^{-1}$ without (NQE) and with the QE approximation listed in table 3. The dominating strong retrapping ($\gamma<0.1$, figure 2c) improves on the QS level at $\omega=1$ $Ks^{-1}$ similarly as for $\omega=10^{-2}$ and $10^{-1} Ks^{-1}$, but the condition for the presence of the QS state ($|q_l| \approx 10^{-2}$) occurs only at high temperatures ($T>210$ K, figure 2d). In the high temperature range ($T=270$-$320$ K) the parameters QE and QS achieve the values $|q_l| \approx 1$ and $|q^*| \approx 1 \times 10^{-2}$ (figure 2d), whereas the recombination probability reaches the value $\gamma \approx 1 \times 10^{-2}$ (figure 2c). These results mean that the rate of change of free electron density is approximately equal to the recombination intensity, but those are much lower that the intensity of retrapping, $|dn_c/dt| \approx I_e << I_t$.

The calculation results of $\gamma(T)$ (figure 2c) and $q_I(T)$ (figure 2d) at $\omega=10^{-2}$ Ks$^{-1}$ demonstrate that the QE state can be realized for the case of strong retrapping. On the other hand, the QE state is not present in the TSL and TSC at $\omega=1$Ks$^{-1}$. However, as it follows from our calculation results, the QE state is reached ($|q_I|<10^{-2}$) at $\omega=1$Ks$^{-1}$ for the set of kinetic parameters (B$_1$) with the densities of the recombination centers and traps that are much higher ($N_t = 10^{15}$ cm$^{-3}$, $M=10^{12}$ cm$^{-3}$) and the other parameters remain as those in the set B. In Case B$_1$ similarly as in Case B the retrapping process is very strong ($\gamma \leq 0.1$) in the whole interesting temperature range. But in Case B$_1$ the QE state appears at $\omega=1$Ks$^{-1}$ because of much lower value of the recombination lifetime ($\tau_e \approx 10^{-2}$ s in the initial range of the TSL intensity $I_e/I_{em}=0.1$) than in Case B ($\tau_e \approx 10$s at $I_e/I_{em}=0.1$). In these two cases the value of $n_c^{-1} dn_c/dT$ is approximately equal to $1.1 \times 10^{-1}$ and, of course, it does not change the value of $q_I$ (see equation (20)).

The results of $q_I(T)$ at $\omega=10^{-2}$ Ks$^{-1}$ for Case B show that the QE level deteriorates at temperatures higher than those at the TSC peak-maximum, $T>T_{mn}=185$ K (figure 2d). This behavior of the $q_I(T)$ dependence can be explained using equation (16) in the form $q_I=G/I-1/\gamma$. It follows from this equation that $q_I>0$ when $G/I_e>1/\gamma$ at $T<T_{mn}$, and $q_I<0$ when $G/I_e<1/\gamma$ at $T>T_{mn}$. At $T>T_{mn}$ $|q_I|$ increases when $\gamma$ decreases (i.e. the retrapping increases) with increasing $T$ (figures 2c and 2d). The temperature dependences of $\gamma$ and $\tau_e$ (figure 2c) at $\omega=1$Ks$^{-1}$ for Case B show similar behavior as for Case A: $\gamma$ decreases slower and $\tau_e$ increases slower with increasing temperature than in the approximate QE state at $\omega=10^{-2}$ Ks$^{-1}$. It suggests that the density of holes in the recombination centres at $\omega=1$Ks$^{-1}$ decreases slower than at $\omega=10^{-2}$ Ks$^{-1}$.

*4.3. Cases C and C$_1$*

The TSL curves calculated without the QE approximation for Case C at $\omega=10^{-2}$ and 1 Ks$^{-1}$ are presented in figure 3a. In this case the recombination probability ($\gamma \approx 0.5$) and the recombination lifetime ($\tau_e \approx 10$ s) are approximately constant in the temperature range of the TSL calculated for the two values of $\omega$. At $\omega=10^{-2}$ Ks$^{-1}$ the TSL curve has the first-order shape with the symmetry factor $\mu_{gI}=0.427$ [2,7,11]. It is interesting to see that the first-order shape can be observed for the case of moderate retrapping ($\gamma \approx 0.5$). The results of $q_I(T)$ calculated for $\omega=10^{-2}$ Ks$^{-1}$ show that there is the approximate QE state ($|q_I|<2.5 \times 10^{-2}$) in the temperature range $T=105-165$ K (figure 3b). In this temperature range the condition for the QS state is better fulfilled ($|q^*|<1.2 \times 10^{-2}$). It results in the well-formed first-order shape of the TSL curve (see equation (18)).

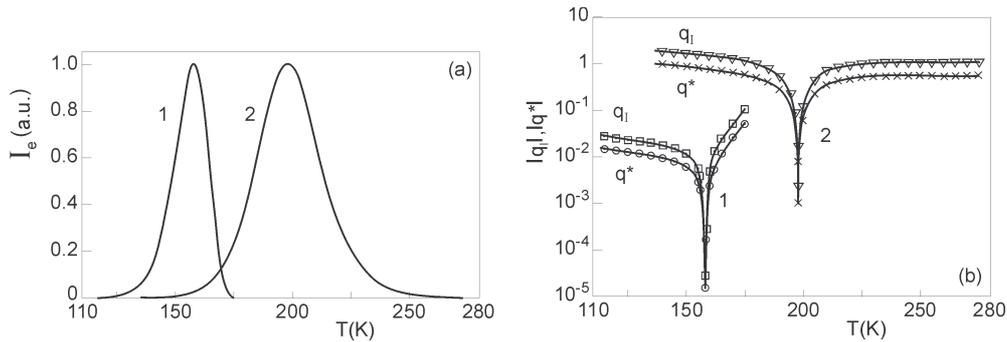

**Figure 3**. Temperature dependences of TSL characteristics calculated for set C of kinetic parameters and for two values of the heating rate ($\omega=10^{-2}$ (1) and 1 (2) Ks$^{-1}$): (a) the relative TSL intensity ($I_e(T)$), and (b) the parameters $q_I(T)$ and $q^*(T)$.

The upper limit value of heating rate for presence of the QS state (with $|q^*|\leq 3.8 \times 10^{-2}$ and $|q_I|\leq 7.2 \times 10^{-2}$) seems to be at $\omega_s \approx 3 \times 10^{-2}$K s$^{-1}$ because the values of the parameters of the NQE and QE TSL curves are in good agreement up to this heating rate value.

The TSL curve calculated for $\omega$=1 Ks$^{-1}$ presents strong deviation in the shape from that for the QE state: it has approximately the second-order shape ($\mu_{gl}\approx 0.5$). In this case neither the QE nor QS states are present (figure 3b).

Similar case (C$_1$) of TSL and TSC is obtained when the values of the density of deep traps and the recombination coefficient are changed in relation to those for Case C (table 1). This is the case (C$_1$) of strong retrapping with very weakly temperature-dependent recombination lifetime and recombination probability ($\tau_e$=5.6x10$^1$s and $\gamma$=1.7x10$^{-1}$ at $I_e(T')/I_{em}$=10$^{-2}$ and $T'<T_{mI}$, and $\tau_e$=6.7x10$^1$s and $\gamma$=1.3x10$^{-1}$ at $I_e(T'')/I_{em}$ =10$^{-2}$ and $T''>T_{mI}$). The parameters of the TSL curves calculated for Case C$_1$ with and without the use of the QE approximation and for three values of the heating rate are presented in table 4.

**Table 4.** Values of parameters of TSL curves calculated for set C$_1$ of kinetic parameters without the QE approximation (NQE) and with the QE approximation (QE) and for three values of heating rate $\omega$

| TSL | $\omega$(Ks$^{-1}$) | $T_{mI}$ (K) | $T_{1I}$(K) | $T_{2I}$(K) | $\mu_{gI}$ | $I_{em}$ (cm$^{-3}$s$^{-1}$) |
|---|---|---|---|---|---|---|
| NQE | 2x10$^{-3}$ | 155.2 | 145.5 | 162.8 | 4.37x10$^{-1}$ | 1.07x10$^6$ |
| QE |  | 155.2 | 145.5 | 162.7 | 4.35x10$^{-1}$ | 1.08 x10$^6$ |
| NQE | 1x10$^{-2}$ | 166.2 | 155.0 | 175.2 | 4.46x10$^{-1}$ | 4.59x10$^6$ |
| QE |  | 166.1 | 155.1 | 174.7 | 4.37x10$^{-1}$ | 4.74x10$^6$ |
| NQE | 10$^{-1}$ | 185.0 | 170.4 | 199.7 | 5.03x10$^{-1}$ | 3.09x10$^7$ |
| QE |  | 184.5 | 170.0 | 195.1 | 4.39x10$^{-1}$ | 3.87x10$^7$ |

The calculation results of the QE parameter for Case C$_1$ ($q_I(T')$=1.5x10$^{-2}$, and $q_I(T'')$=2.7x10$^{-2}$) suggest that the heating rate $\omega_q$=2x10$^{-3}$ Ks$^{-1}$ corresponds to the upper limit of presence of the QE state. The upper limit of $\omega$ for the QS state ($q^*(T')$=1.0x10$^{-2}$, and $q^*(T'')$=1.5x10$^{-2}$) seems to be at $\omega_s\approx$10$^{-2}$ Ks$^{-1}$. At this heating rate the shape of the NQE TSL curve ($\mu_g$=0.446) is still similar to the first order shape ($\mu_g$=0.424) in spite of the strong retrapping. At higher heating rates (e.g. $\omega$=10$^{-1}$Ks$^{-1}$) the shape of the NQE TSL curve ($\mu_{gI}$=0.503) becomes similar to the second order shape. The TSC curves calculated for Cases C and C$_1$ and for $\omega$=10$^{-2}$ and 1Ks$^{-1}$ have approximately the same shape as the corresponding TSL curves because of the nearly constant recombination lifetime during the TSL and TSC processes in these cases.

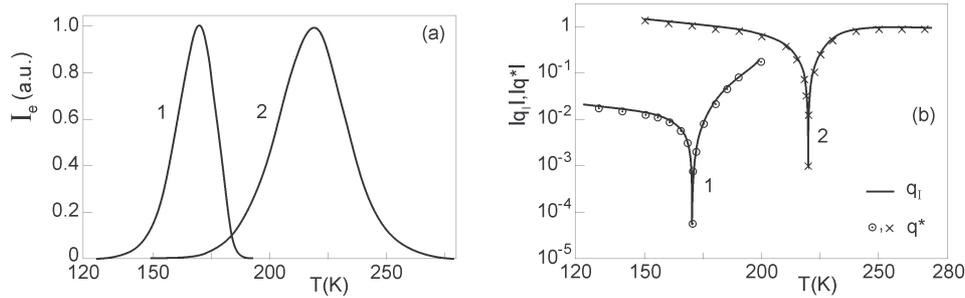

**Figure 4**. Temperature dependences of TSL and TSC characteristics calculated for set D of kinetic parameters and for two values of the heating rate ($\omega$=10$^{-2}$ (1) and 1 (2) Ks$^{-1}$): (a) the relative TSL intensity ($I_e(T)$), and (b) the parameters $q_I(T)$ and $q^*(T)$.

*4.4. Case D*

Figure 4a presents the TSL curves calculated for set D of kinetic parameters and $\omega$=10$^{-2}$ Ks$^{-1}$. In this case the recombination lifetime very weakly increases with increasing temperature from $\tau_e$=9s to 10s within the temperature range of the TSL curve. The recombination probability shows even weaker temperature dependence and its value is typical for the weak retrapping ($\gamma\approx$0.9). The TSL and TSC curves calculated for $\omega$=10$^{-2}$ Ks$^{-1}$ have the first-order shape with the symmetry factor $\mu_g$=0.426. Figure 4b shows that the TSL and TSC processes at $\omega$=10$^{-2}$Ks$^{-1}$ are realized under approximately fulfilled QE condition ($|q_I|$<2x10$^{-2}$ at $T$=130-177 K ). Because of the value of $\gamma\approx$0.9 the values of the parameters $q_I$ and $q^*$ differ only by about 10%. At $\omega$=1 Ks$^{-1}$ the QE state and both the QS state are not realized. Consequently the shapes of the TSL and TSC curves (with $\mu_g$=0.473, figure 4a) differ from the typical first-order kinetics shape.

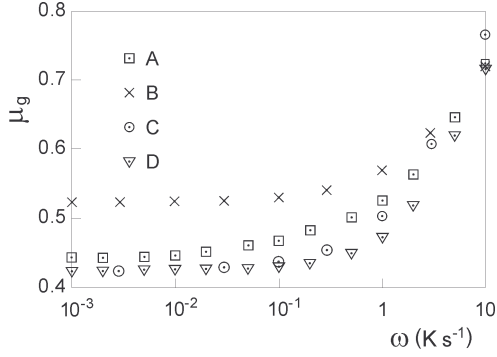

**Figure 5**. Symmetry factor of TSL curve ($\mu_g$) as a function of heating rate ($\omega$) calculated for sets A, B, C, and D of kinetic parameters.

## 5. Determination of the QE and QS parameters

A proposition of the method for estimation of the QE level was given for the first time by Sunta and colleagues [24]. The method is based on the dependence of the symmetry factor of the TSL curve on the heating rate, $\mu_{gI}(\omega)$. According to the method, the upper limit value of $\omega_q$ for presence of the QE state can be determined from the turning point of the $\mu_{gI}(\omega)$ curve from the weak to the strong dependence of $\mu_{gI}$ on $\omega$. Figure 5 shows the dependence of $\mu_{gI}$ on $\omega$ for the TSL curves calculated for the above described Cases A,B,C, and D. It is seen from figure 5 that the turning point on the $\mu_{gI}(\omega)$ curves for the weak retrapping Case A appears approximately at $\omega_q$=10$^{-2}$ which corresponds to the end of presence of the QE and QS states determined by $|q_I|\approx |q^*|\approx$10$^{-2}$ (Section 4). For the weak retrapping Case D the turning point on the $\mu_g(\omega)$ curve is seen at $\omega\approx$10$^{-1}$Ks$^{-1}$ whereas the upper limit for presence of the QE and QS states is at $\omega_q\approx\omega_s\approx$10$^{-2}$ (Section 4). For the moderate and strong retrapping cases (C, and B and C$_1$) the turning point appears at higher values of $\omega$ than the upper limit of presence of the QE state. The results of $q^*(\omega)$ for the strong retrapping case B show that the turning point on the dependence $\mu_g(\omega)$ is at the limit of the presence of the QS state that appears at $\omega_s\approx$10$^{-1}$Ks$^{-1}$ whereas the limit for the QE state is at $\omega_q\approx$10$^{-2}$Ks$^{-1}$ (Section 4). Similarly, the turning point for Case C is seen at $\omega_s\approx$3x10$^{-2}$ Ks$^{-1}$ but the QE limit is at $\omega_q\approx$10$^{-2}$ Ks$^{-1}$ (Section 4). The QE state for Case C$_1$ is present at lower heating rates ($\omega_q\leq$2x10$^{-3}$Ks$^{-1}$) than the turning point that is at $\omega_s\approx$10$^{-2}$ Ks$^{-1}$.

We analyzed also the dependences of the peak-maximum intensity of the TSL and TSC on the heating rate and the reciprocal peak-maximum temperature.

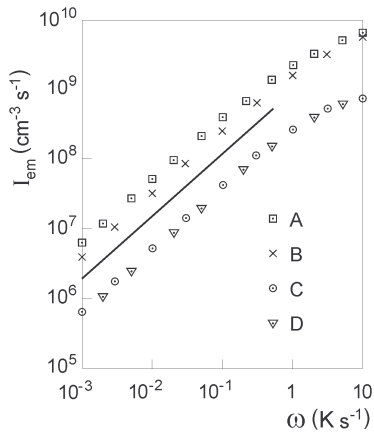

**Figure 6**. Peak-maximum intensity of TSL curve ($I_{em}$) as function of heating rate ($\omega$) calculated for Cases A, B, C, and D. The straight line presents the dependence $I_{em}\sim\omega^{0.9}$.

Analysis of the $I_{em}(\omega)$ curves shows that the approximate linear dependence of $I_{em}$ on $\omega$ ($I_{em} \propto \omega^k$ with $k \approx 0.9$, figure 6) is observed in wide range of $\omega$ in all four Cases A,B,C, and D. Similar dependence of the peak-maximum intensity of the TSC on the heating rate ($n_{cm} \propto \omega^k$ with $k \approx 0.9$) is valid in these cases. The values of the factor k becomes lower ($k<0.9$) at the high heating rates (figure 6). Analysis of the results of $I_{em}(\omega)$, $q_I(\omega)$, and $q^*(\omega)$ shows that the dependence of $I_{em}$ on $\omega^{0.9}$ is present in the range of $\omega$ corresponding to the QS state. For example, the QS state of the TSL peak-maximum intensity in Case B occurs up to $\omega_s = 5 \times 10^{-1}$ Ks$^{-1}$ whereas the upper limit of $\omega$ for the QE state is at $\omega_q \approx 10^{-2}$ Ks$^{-1}$ (figures 2a and 2c).

Figure 7 presents the dependence of the TSL peak-maximum intensity $I_{em}$ on the related reciprocal temperature $T_{mI}^{-1}$ calculated for Cases A,B,C and D and for different heating rates $\omega_i$. It is seen from the figure that for the weak retrapping cases (A and D) the functions of $I_{em}(1/T_{mI})$ have exponential character in the range of the validity of the QE and QS states ($\omega_q \approx \omega_s \approx 10^{-2}$Ks$^{-1}$). The values of $q_{Ii}$ and $q_i^*$ corresponding to $I_{emi}$ calculated for different values of $\omega_i$ for Cases B and C (see figures 2c and 3b and Sections 4.2 and 4.3) show that the limit of presence of the QS state is at higher value of $T_{mI}$ than that at the QE state. It causes that the exponential dependence of $I_{em}$ on $T_{mI}^{-1}$ in Cases B and C is observed up to the limit of presence of the QS state ($\omega_s = 10^{-1}$Ks$^{-1}$ for Case B, and $\omega_s = 3 \times 10^{-2}$Ks$^{-1}$ for Case C). At the values of $\omega_i$ higher than $\omega_s$ (i.e. beyond the QS state) one observes strong deviation of the $I_{emi}(1/T_{mIi})$ function from the exponential one (figure 7).

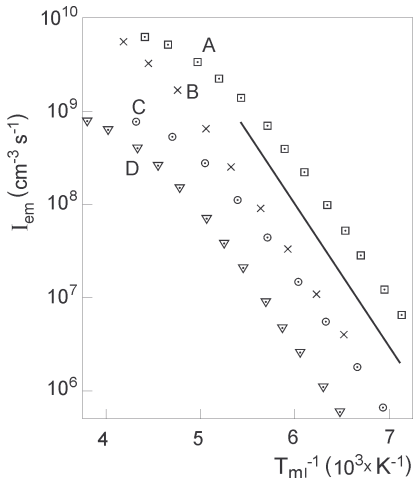

**Figure 7**. Peak-maximum intensity of TSL curve ($I_{em}$) as function of reciprocal peak-maximum temperature ($T_{mI}^{-1}$) calculated for Cases A, B, C, and D and for different heating rates. The straight line shows the slope -0.3 eV/k.

The dependences of the TSC peak-maximum intensity ($n_{cmi}$) on the reciprocal temperature ($T_{mni}^{-1}$) calculated for different values of $\omega_i$ for Cases A, B, C, and D have character similar to the corresponding dependences of $I_{emi}$ on $T_{mIi}^{-1}$. They are exponential in the range of occurrence of the QS state.

In Section 2 we have derived simple formula for calculation of the ratio $q_{I2}/q_{I1}$ (27), where $q_{I1}$ and $q_{I2}$ are the QE parameters for equal value of $I_e/I_{em}$ (or $n_c/n_{cm}$) in the initial range of the TSL (TSC) at two values of the heating rate $\omega_1$ and, respectively, $\omega_2$. The formula was derived under the assumption that the parameters $\tau_e$, $R$, and $q^*$ are independent of the temperature in the initial part of the TSL or TSC curve and that $\tau_e$ corresponding to the fixed value of $I_{ein}/I_{em}$ is independent of the heating rate.

We therefore calculated the values of $\tau_e$, $R$ and $q^*$ for sets A,B,C, and D of kinetic parameters and different values of the relative TSL intensity ($I_e/I_{em}$) and different values of the heating rate($\omega$). The resulting values of $R, \tau_e$, and $q^*$ calculated for $I_e/I_{em} = 5 \times 10^{-2}$ and $3 \times 10^{-1}$ (at $T<T_{mI}$), and $\omega = 10^{-2}$ and 1 Ks$^{-1}$ are given in table 5. The values of $R$, $\tau_e$, and $q^*$ show that in the TSL initial range these parameters are weakly dependent on the temperature. The parameters $R$ and $\tau_e$ are also very weakly dependent on the heating rate.

**Table 5**. Values of parameters $R$, $\tau_e$, and $q^*$ calculated for sets A, B, C, $C_1$, and D of kinetic parameters and three values of relative TSL intensity ($I_e/I_{em}$) in its initial range, and for two values of heating rate $\omega$.

| | $I_e/I_{em}$ | 5x10$^{-2}$ | | | 3x10$^{-1}$ | | |
|---|---|---|---|---|---|---|---|
| Case | $\omega$(Ks$^{-1}$) | $R$ | $\tau_e$(s) | $q^*$ | $R$ | $\tau_e$(s) | $q^*$ |
| A | 10$^{-2}$ | 9.86x10$^{-1}$ | 1.01x10$^{1}$ | 2.05x10$^{-2}$ | 9.01x10$^{-1}$ | 1.09x10$^{1}$ | 1.88x10$^{-2}$ |
|   | 1 | 9.79x10$^{-1}$ | 1.00x10$^{1}$ | 1.45 | 8.64x10$^{-1}$ | 1.06x10$^{1}$ | 1.19 |
| B | 10$^{-2}$ | 9.91x10$^{-3}$ | 9.99 | 1.75x10$^{-2}$ | 8.82x10$^{-3}$ | 1.06x10$^{1}$ | 1.40x10$^{-2}$ |
|   | 1 | 9.88x10$^{-3}$ | 9.96 | 1.24x10$^{-1}$ | 8.65x10$^{-3}$ | 1.04x10$^{1}$ | 9.28x10$^{-2}$ |
| C | 10$^{-2}$ | 5.21x10$^{-2}$ | 9.90 | 9.84x1 | 4.74x10$^{-2}$ | 9.91 | 7.72x10$^{-3}$ |
|   | 1 | 5.16x10$^{-2}$ | 9.90 | 7.08x10$^{-1}$ | 4.43x10$^{-2}$ | 9.91 | 5.22x10$^{-1}$ |
| D | 10$^{-2}$ | 9.12x10$^{-2}$ | 9.10 | 1.41x10$^{-2}$ | 8.34x10$^{-2}$ | 9.17 | 1.11x10$^{-2}$ |
|   | 1 | 9.04x10$^{-2}$ | 9.10 | 8.63x10$^{-1}$ | 7.84x10$^{-2}$ | 9.16 | 7.22x10$^{-1}$ |

**Table 6**. Values of ratio of parameters $q_{I2}=q_I(\omega_2)$ and $q_{I1}=q_I(\omega_1)$ calculated for sets A, B, C, $C_1$, and D and for two values of the heating rate $\omega_1$=10$^{-2}$Ks$^{-1}$ and $\omega_2$=1Ks$^{-1}$ and for two values of relative TSL intensity $I_e/I_{em}$ using approximate formula (28). Values of $q_{I0}$ and $q_{I20}$ are calculated using equation (9) and values of $dn_c/dt$ and $I_e$ calculated by solution of the kinetic equations system (1-4).

| $I_e/I_{em}$ | 1.5x10$^{-1}$ | | 3x10$^{-1}$ | |
|---|---|---|---|---|
| Case | $q_{I20}/q_{I10}$ | $q_{I2}/q_{I1}$ | $q_{I20}/q_{I10}$ | $q_{I2}/q_{I1}$ |
| A | 6.76x10$^{1}$ | 6.72x10$^{1}$ | 6.34x10$^{1}$ | 6.43x10$^{1}$ |
| B | 6.81x10$^{1}$ | 6.90x10$^{1}$ | 6.54x10$^{1}$ | 6.80x10$^{1}$ |
| C | 6.95x10$^{1}$ | 6.75x10$^{1}$ | 6.78x10$^{1}$ | 6.67x10$^{1}$ |
| $C_1$ | 8.04x10$^{1}$ | 6.98x101 | 6.79x10$^{1}$ | 8.29x10$^{-1}$ |
| D | 6.68x10$^{1}$ | 6.42x10$^{1}$ | 6.54x10$^{1}$ | 6.33x10$^{1}$ |

Table 6 presents the values of the ratio $q_{I2}/q_{I1}$ determined with the above mentioned method (equation (28)). They are in good agreement with the corresponding results of $q_{I20}/q_{I10}$ calculated using equation (9).

In this work we have also proposed the method for determination of two values of the QS parameter, $q_1^*$ and $q_2^*$, from the initial parts of two TSL curves measured at two different heating rates, $\omega_1$ and $\omega_2$ (Section 2). The method uses equations (32) and (33) that are derived under the assumptions that the parameters $\tau_e$ and $R$ corresponding to equal values of the relative TSL intensity (in the TSL initial range) are independent of the value of $\omega$.

**Table 7**. Values of QS parameter calculated for the sets A,B,C, $C_1$, and D, and for the relative initial TSL intensity $I_e/I_{em}$=1.5x10$^{-1}$, and for two values of heating rate $\omega$. Values of $q^*$ are calculated using the "two equations system" ((32) and (33)) and values of $I_e(T_i)$ and $T_i$ (i=1,2) for $\omega_1$=10$^{-1}$ Ks$^{-1}$ and $\omega_2$=1 Ks$^{-1}$, and E=0.3 eV. Values of $q_0^*$ are calculated using equation (23) and values of $dn_c/dt$ and $I_e$ found by solution of kinetic equations system (1-4).

| $\omega$(Ks$^{-1}$) | 10$^{-1}$ | | 1 | |
|---|---|---|---|---|
| Case | $q_{01}^*$ | $q_1^*$ | $q_{02}^*$ | $q_2^*$ |
| A | 1.61x10$^{-1}$ | 1.74x10$^{-1}$ | 1.30 | 1.37 |
| B | 1.30x10$^{-2}$ | 1.45x10$^{-2}$ | 1.07x10$^{-1}$ | 1.20x10$^{-1}$ |
| C | 8.10x10$^{-2}$ | 8.28x10$^{-2}$ | 6.10x10$^{-1}$ | 6.70x10$^{-1}$ |
| $C_1$ | 1.20x10$^{-1}$ | 1.40x10$^{-1}$ | 1.02 | 1.16 |
| D | 1.03x10$^{-1}$ | 1.15x10$^{-1}$ | 8.34x10$^{-1}$ | 9.01x10$^{-1}$ |

Table 7 presents the values of $q_1^*$ and $q_2^*$ determined by applying the "two equations method" to the initial range of the TSL curves calculated for sets A,B,C,$C_1$, and D of kinetic parameters and two values of the heating rate. These results of $q^*$ are compared with the corresponding results, $q_0^*$, calculated using formula (23). Relative error of the values of $q^*$ determined with the "two equations method" is lower than 17% when the method is applied to the relative TSL intensity of $I_e/I_{em}$=0.15, and the error is not higher than 50% when $I_e/I_{em}$=0.3.

Similar method for determination of the value of $q^*$ is also based on the analysis of the initial parts of the TSL curves measured at two different heating rates. The method uses equation (36) that is derived from equation (31) under the assumption that the slower heating rate is low sufficiently to obtain the QS state ($\omega_1=\omega_s$). The values of $q^*$ were calculated by applying equation (36) to the initial parts of the TSL curves computed for the sets A, B, C, $C_1$, and D of kinetic parameters and two values of the heating rate. On the basis of the results presented in this section and in Section 4 we assumed that the TSL generated at $\omega_1=10^{-2}$ Ks$^{-1}$ is in the QS state. The results of the determination of $q^*$ are presented in table 8. It is seen from the table that the above-described method provides the values of $q^*$ with the accuracy better than 11% when evaluated for $I_e/I_{em}$=0.15, and with lower accuracy (<25%) for $I_e/I_{em}$=0.3.

**Table 8**. Values of QS parameter calculated for the sets A, B, C, $C_1$, and D and for two values of relative initial TSL intensity $I_e/I_{em}$ and heating rate $\omega$=1Ks$^{-1}$. Values of q* are calculated using approximate formula (36) and values of $I_e(T_i)$ and $T_i$ (I=1,2) for $\omega_1=\omega_s= 10^{-2}$Ks$^{-1}$ and $\omega_2$ =1Ks$^{-1}$. Values of $q_0^*$ are calculated using Eq.(23).

| $I_e/I_{em}$ | 1.5x10$^{-1}$ | | 3x10$^{-1}$ | |
|---|---|---|---|---|
| Case | $q_0^*$ | $q^*$ | $q_0^*$ | $q^*$ |
| A | 1.30 | 1.30 | 1.19 | 1.24 |
| B | 1.07x10$^{-1}$ | 1.16x10$^{-1}$ | 9.28x10$^{-2}$ | 1.12x10$^{-1}$ |
| C | 6.10x10$^{-1}$ | 6.48x10$^{-1}$ | 5.22x10$^{-1}$ | 6.16x10$^{-1}$ |
| $C_1$ | 1.02 | 1.13 | 8.50x10$^{-1}$ | 1.10 |
| D | 8.34x10$^{-1}$ | 8.63x10$^{-1}$ | 7.23x10$^{-1}$ | 8.10x10$^{-1}$ |

## 6. Estimation of the accuracy of the determination of the trap depth by the literature methods

As it is above described the value of the active trap depth E is necessary for the determination of the QS parameter q*. We therefore determined the trap depth from the TSL and TSC curves calculated without QE (non-QE) approximation for Sets A,B, C, and D of kinetic parameters (table 1). The trap depth (E) values were determined using the methods well-known from literature: the initial-rise method of Garlick and Gibson [4], the heating-rate methods of Hoogenstraaten [5], and Haering and Adams [6], and the peak-shape methods of Chen [2]. All the methods yield the values of E with the relative error lower than 8% when applied to the QS TSL curves calculated at low heating rates, $\omega \leq 10^{-2}$ Ks$^{-1}$. These results are in agreement with the previously reported results found for the TSL curves calculated using the QE approximation [10,28]. These methods applied to the non-QS TSL and TSC curves calculated at the higher heating rate ($\omega$=1Ks$^{-1}$) generate the values of E with different accuracy. The method of Garlick and Gibson applied to the non-QS TSL and TSC curves generates the values of E with small error (4 %). It results from the fact that the exponential factor (exp[-E/kT]) is dominating in the temperature dependence of the TSL intensity in the initial part of the TSL curve also at the high heating rate. The peak-shape $\tau$-method of Chen applied to the non-QS TSL curves yields also the values of E with reasonable accuracy (10%). This is probably because the shape of the low temperature part of the non-QS TSL curve is weakly changed from the shape of the QE TSL curve. On the other hand the $\delta$-method of Chen and the two heating-rate methods can produce very large errors of the values of E (up to 40%). The reason of the size of these errors of the values of E is strong deviation of the parameters (the peak position, the intensity, and the shape) of the non-QS TSL (TSC) curve from the parameters of the corresponding QE TSL (TSC) curve.

## 7. Conclusion

In this work we have considered the TSL and TSC processes using the classical kinetic model of n-type insulator with two kinds of the electron traps (the active (shallow), and the non-active (deep-thermally disconnected)), and one kind of the recombination centres. The kinetic equations describing the model are solved without and with the quasi-equilibrium (QE) approximation. The TSL and TSC processes are analyzed using the QE parameter $q_I$ and a new parameter $q^*=\gamma q_I$ describing the relation between the rate of change of free electrons density and the intensities of recombination and trapping ($\gamma \leq 1$ is the relative recombination probability). The QE state and the quasi-stationary (QS) state are determined by $|q_I|<<1$, and, respectively, $|q^*|<<1$. Approximate formulas for determination of the relative QE parameter $q_{I2}/q_{I1}$, and the QS parameter $q^*$ from the initial range of the TSL or TSC are derived. The methods are based on analysis of two TSL (or TSC) curves measured at two different heating rates ($\omega_1$ and $\omega_2$). We can find the value of $q_{I2}(T_2)/q_{I1}(T_1)$ using equation (28) and knowing values of the temperatures $T_1(\omega_1)$ and $T_2(\omega_2)$ corresponding to equal value of the relative TSL (TSC) intensity in its initial rage. Two values of the QS parameter ($q_1^*$ and $q_2^*$) can be determined by applying the "two equations method" (equations (32) an (33)) to the initial parts of two TSL curves measured at two different heating rates. The parameters of the TSL curve that are needed for determination of the values of $q_1^*$ and $q_2^*$ can be evaluated from experiment: (1) two values of the TSL intensity (in relative units) measured at two different heating rates and corresponding to equal values of the relative TSL intensity, (2) the values of the heating rate, and (3) the value of the trap depth. Another method (equation (36)) for the determination of the value of $q^*$ (a variant of the "two equation method") for the TSL in the non-QS state measured at some heating rate ($\omega_2$) is effective when the second TSL curve is measured at heating rate ($\omega_1$) slow enough to obtain the QS state.

The TSL and TSC curves and other characteristics ( the QE and QS parameters, the recombination lifetime, the recombination probability, the density of electrons trapped in the active traps, the densities of holes captured by the recombination centers) are calculated for five sets of kinetic parameters and different heating rates. The characteristics calculated at low heating rates have the properties typical for the QE state. The properties of the TSL and TSC characteristics calculated at high heating rates deviate from those typical for the QE state. The deviation of the properties appears when the heating rate is higher than that corresponding to the limit of validity of the QS state.

In the case of the weak retrapping ($\gamma \approx 1$) the QE and QS states appear in approximately the same range of the heating rate i.e. the upper limit of heating rate is approximately the same for the presence of both the QE and QS state. However, in the case of the non-negligible retrapping ($\gamma \ll 1$) the limit of the validity of the QS state appears at higher heating rate than the highest heating rate permitting the presence of the QE state.

The QE state of the TSL and TSC can appear both in the weak and strong retrapping processes. It means that the strong retrapping is not in contradiction with the QE state. It is so because the value of the recombination lifetime is fundamental for reaching of the QE state. The QS TSL curves for the case of strong retrapping have the so-called first-order shape (typical for the weak retrapping) when the recombination lifetime is independent of the temperature. It may contribute to dominating presence of the first-order shape of the TSL curve in nature. These conclusions concerning the influence of the retrapping process on the TSL and TSC properties are in agreement with the corresponding conclusions in the earlier works [19,20,22-24].

Our results suggest that the range of the heating rate of a crystal allowing occurrence of the QS state of TSL and TSC processes can be approximately estimated from the dependences: (1) very weak increase of the symmetry factor of the TSL (TSC) curve with increasing heating rate (2) approximately linear dependence of the TSL (TSC) peak-maximum intensity on the heating rate, and (3) exponential dependence of the TSL (TSC) peak-maximum intensity on the reciprocal peak-maximum temperature.

The above-mentioned methods for determination of the parameters $q_{I2}/q_{I1}$, and $q^*$ in the initial range of TSL and TSC were examined by applying them to the TSL and TSC curves calculated without the use of the QE approximation for five sets of kinetic parameters. The results of the examination show that the methods yield reasonable values of the parameters (error <30%) when applied to the initial range of the TSL (or TSC) with the intensity not higher than 30% of the maximum intensity. It should be noticed that in the case of the weak retrapping ($\gamma \approx 1$) the methods for determination of the QS parameter $q^*$ yield also the approximate value of the QE parameter $q_I$ because of the relation $q^*=q\gamma_t$. Having the vales of the initial parameter $q^*$ one can roughly estimate its values decreasing with the increasing temperature up to the TSC peak-maximum position where $q_I=q^*=0$.

The calculated TSL and TSC curves were also used to examine the accuracy of four methods (Hoogenstraaten, Haering and Adams, Garlick and Gibson, and Chen) for determination of the trap depth. The examination of the methods shows that only the initial-rise method of Garlick and Gibson produces accurate values of the trap depth (error <5%) when applied to the non-QE TSL and TSC curves. The results of estimation of the accuracy of the methods of Hoogenstraaten and Garlick and Gibson applied to the TSL curves are in agreement with the results of Lewandowski and McKeever [14].

## Acknowledgment

Many thanks are due to Dr. A. Brozi from our institute for reading the manuscript of this article and critical comments.